\documentclass[preprint,showpacs,preprintnumbers,amsmath,amssymb]{revtex4}

\usepackage{graphicx}
\usepackage{amsmath}
\usepackage{amssymb}
\usepackage{bm}
\usepackage{mathrsfs}
\usepackage{color}
\usepackage[normalem]{ulem}
    \usepackage{braket}

\newcommand{\oad}{\hat{a}^\dagger}
\newcommand{\oa}{\hat{a}}

\newcommand{\oald}{\hat{\alpha}^\dagger}
\newcommand{\oal}{\hat{\alpha}}

\begin{document}

\title{Superconducting qubits beyond the dispersive regime}
\author{M.H. Ansari}
\affiliation{ Peter Gr\"unberg Institute, Forschungszentrum J\"ulich, Germany}
\affiliation{ J\"ulich-Aachen Research Alliance (JARA), Fundamentals of Future Information Technologies, Germany}

\begin{abstract}
Superconducting circuits consisting of a few low-anharmonic transmons coupled to readout and bus resonators can perform basic quantum computations. Since the number of qubits in such circuits is limited to not more than a few tens, the qubits can be designed to operate within the dispersive regime, where frequency detuning are much stronger than coupling strengths.  However, scaling up the number of qubits will bring the circuit out of this regime  and   invalidates current theories. We develop a formalism that allows to consistently diagonalize superconducting circuit hamiltonian beyond dispersive regime. This will allow to study qubit-qubit interaction unperturbatively, therefore our formalism remains valid and accurate at small or even negligible frequency detuning; thus our formalism serves as a theoretical ground for designing qubit characteristics for scaling up the number of qubits in  superconducting circuits. We study the most important circuits with single- and two-qubit gates, i.e. a single transmon coupled to a resonator and two transmons sharing a bus resonator. Surprisingly our formalism allows to determine the circuit characteristics, such as dressed frequencies and Kerr couplings,  in closed-form formulas that not only reproduce perturbative results but also extrapolate beyond the dispersive regime and can ultimately reproduce (and even modify) the Jaynes-Cumming results at resonant frequencies. 
\end{abstract}


\maketitle

\section{Introduction}

Quantum computation is rapidly progressing toward practical technology \cite{{Lucero},{Kandala17},{Cai},{Monz}}. So far quantum bits have been well developed on superconducting circuits \cite{Wilhelm}. When cooled to milikelvin temperatures, coherent tunneling of the Cooper pairs through the Josephson junction (JJ) exhibits slightly nonlinear harmonic oscillations with addressable energy levels \cite{Gambetta16}. Such quantum states have long coherent times and can operate in nanosecond scales. Moreover, they are compatible with the well-established microwave control technology and can scale up in large numbers. All these features makes superconducting qubits one of the prominent platforms for constructing a multiqubit quantum processor  \cite{{Chow},{Niemczyk}}. 

The state of art superconducting circuits contains a few tens of qubits with operational gate error rate about 0.1$\%$ for single qubit gates \cite{Barends}  and 1$\%$ for two-qubit gates \cite{Sheldon} yet marginally below the threshold for error detection in the surface code \cite{Gambetta17}. Scaling up the qubit number to be more than a few tens will dramatically increases errors to unacceptable values and the key milestone of next few years is to reduce the errors \cite{{Preskill},{Martinis18}}. Achieving this not only requires further enhancements in the circuit quality \cite{Martinins16} but also needs progressive advancement in theory \cite{theory}. So far the Jaynes-Cummings model, originally introduced in quantum optics \cite{JC}, and its generalization have been routinely applied on superconducting circuits. These models have been so far well-studied for parameters admissible by perturbation theory, namely within `dispersive regime' \cite{{Wallraff11},{Nataf}}. Moreover, special resonant frequency solutions are also known \cite{{Haroche},{Blais04}}.  However, scaling up the number of qubits within the narrow domain of parameters will introduce new issues, such as circuit frequency crowding that must be avoided \cite{Plourde17}. Recently it has been discussed that significant advantages can be made in engineering circuits outside of the dispersive regime \cite{Bultkin}. However, those studies have been performed numerically in the absence of established theory. 

\begin{figure}[t]
\vspace{-0.5cm}
		\includegraphics[width=0.7\linewidth]{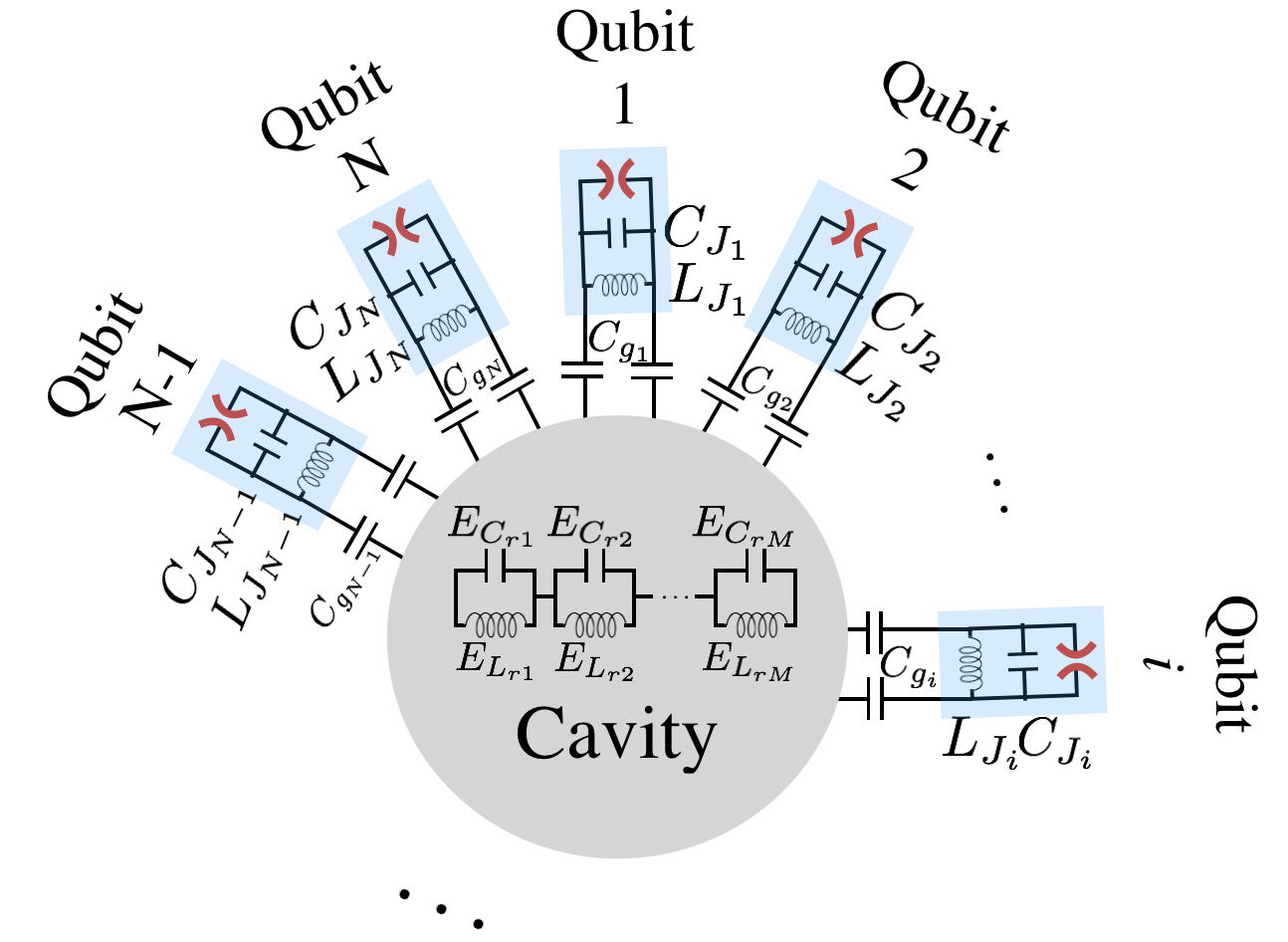}
		\vspace{-0.4cm} \caption{$N$ transmons (blue boxes)  coupled to a multimode resonator; the curly (red) crosses are the nonlinear JJ's that represent the anharmonicity sector, everything else makes the harmonic one.}
		\label{fig schem}
	\end{figure}

Here, motivated by the `black box quantization' method,  we develop a formalism for evaluating the qubit characteristics in circuits consisting of transmons and multimode resonators/cavities at arbitrary frequencies and coupling strengths. Black box quantization has been recently introduced in Ref. \cite{Nigg12} for circuits consisting of low anharmonic transmons coupled to resonators. The low anharmonicity allows dividing the circuit hamiltonian into harmonic and anharmonic sectors. In the absence of anharmonicity the transmons and the resonators can be treated on equal footing, thus the Foster decomposition \cite{Foster} can replace the harmonic circuit with a set of lumped imaginary impedances seen by the anharmonic sector. Identifying the characteristic impedances is of the central importance in this method for which Ref. \cite{Nigg12} proposes iterative feedbacks between experiment and theory.    This formalism has initiated so far several progressive improvements for extracting circuit parameters from electromagnetic simulation \cite{{tureci17},{Firat17},{Richer17}}.  In Fig.(\ref{fig schem})  the harmonic sector is made of $N$ qubits (in blue boxes)  coupled to cavity modes (in the gray area). The curly (red) crosses denote anharmonic sector. 

After introducing a unitary transformation matrix in the space of total number of qubits and resonators, we find a normal-mode basis for the harmonic sector of a multiqubit circuit.  Using the transformation we determine all dressed frequencies and Kerr nonlinear terms in the leading order of anharmonicity.   The simplicity and accuracy of this method allows us to present he results of a transmon and two transmons in closed form formulas. For complex circuits this method provides insightful hamiltonian diagonalization inside and outside of the dispersive regime, which will be progressively useful in scaling up the number of qubits. We explain the formalism first in single transmon, then we generalize it before we solve another example of two transmons.

\section{Single transmon coupled to resonator}

\emph{A transmon coupled to a resonator --} The canonical variables are charges and phases \cite{Clerk}, i.e. $({q}_i, {\phi}_i)$  with $i$ being $t, r$ for transmon  and resonator.  The transmon is  coupled to the center conductor of resonator by the capacitance  $C_g$. The dipole interaction $H_{int}=\beta V_r q_t$ couples the transmon charge and the resonator voltage $V_r=q_r/C_r$,  with  $C_{r/t}$  being the resonator/transmon capacitance and $\beta\equiv C_g/C_t$. Keeping $\beta \ll 1 $  guarantees the increase in qubit coherence time \cite{Manucharyan}.  The circuit harmonic and anharmonic sectors sum to define the circuit classical hamiltonian:   
\begin{eqnarray} \nonumber
H=H_{\textup{har}}+H_{\textup{anhar}}, \ \ \ \ H_{\textup{anhar}}=- \frac{E_C }{3 Z_{t}^2 \hbar^2}  \phi_t^4 && \\ 
     H_{\textup{har}}=\sum_{i=r,t} \frac{q_i^2}{2C_i} + \frac{\phi_i^2}{2L_i}  +H_{int} &&   \label{eq. Hexm1}\end{eqnarray}
 
 The characteristic impedances and the harmonic  frequencies in the circuit are $Z_{i}=\sqrt{L_i/C_i}$ and  $\omega_{i}=1/\sqrt{L_i C_i}$, respectively, with $E_C$ being total capacitive energy of transmon (including JJ and shunt capacitances as well as  capacitive coupling  between transmon and voltage sources),  and $\hbar$ the reduced Planck constant.   We define  canonical variables $({Q}_i, {X}_i)\equiv ({q}_i \sqrt{L_i},{\phi}_i/ \sqrt{L_i})$ such that the harmonic part of Eq. (\ref{eq. Hexm1})  can be transformed to  
\begin{equation}
H_{\textup{har}}=\frac{1}{2}\mathbf{Q}^\textup{T} \mathbf{M Q}+\frac{1}{2}\mathbf{X}^\textup{T}\mathbf{  X}; \  \mathbf{M}=\left[\begin{array}{ccccc}
\omega_{t}^{2} &  g \sqrt{4\omega_{t}\omega_{r}}\\
g \sqrt{4\omega_{t}\omega_{r}} & \omega_{r}^{2}  \label{eq:M}
\end{array}\right],
\end{equation}
with $g \equiv \beta \omega_r \sqrt{Z_r/4Z_t}$ and $\mathbf{Q}\equiv (Q_t,Q_r)$ and  $\mathbf{X}\equiv (X_t,X_r)$.

This Hamiltonian can be diagonalized by unitarily transforming $\mathbf{Q}$ and $\mathbf{X}$ into new canonical variable $\mathcal{Q}$ and $\mathcal{X}$, i.e.  ${Q}_i=\sum_j S_{ij} \mathcal{Q}_j$ and ${X}_i=\sum_j T_{ij} \mathcal{X}_j$.  Given that the variables in  the new and the old frames must satisfy the Poisson brackets of canonical coordinates, i.e. $\{\mathcal{Q}_i,\mathcal{X}_j\}=\{Q_i,X_j\}= \delta_{ij}$,  one can  find  that $T_{ij}=S_{ij}$   (see Appendix \ref{sec.Unitary_SM}).  The only term in Eq. (\ref{eq:M}) that needs diagonalization is  $\mathbf{Q}^\textup{T}\mathbf{ M Q}$, which in the new basis must look like  $\mathcal{Q}^\textup{T} \Omega \mathcal{Q}$,  with $\Omega$ being a diagonal matrix,  $\Omega_{tt}=\bar{\omega}_t^2$, $\Omega_{rr}=\bar{\omega}_r^2$, and zero otherwise.  The unitary transformation  $\mathbf{S}$ is therefore the matrix of columns of normalized eigenvectors of $\mathbf{M}$.  

In the new basis the following dressed frequencies can be found in the linear sector: $\bar{\omega}_t\equiv K_-^2$ and $\bar{\omega}_r\equiv K_+^2$,  with $K_\pm \equiv 2^{-\frac{1}{4}} \left( \omega_t^2+\omega_r^2\pm \Delta \Sigma s^{-1}\right)^{\frac{1}{4}}$ and $s\equiv ({1+{16 \left({g}/{\Delta}\right)^2 \omega_r \omega_t}/{\Sigma^2} })^{-1/2}$,  $\Sigma\equiv \omega_r+\omega_t$ and $\Delta\equiv \omega_r-\omega_t$.  The unitary transformation matrix $\mathbf{S}$  is made of columns of the following normalized eigenvectors $(\pm \sqrt{({1\mp s})/2}, \sqrt{({1\pm s})/2})^T$ associate to the eigenvalues $K_{\pm}^2$.  (In  Appendix \ref{sec.Bogoliuboc_SM} similar results have been found using the Bogoliubov transformations \cite{BV}) In this basis the anharmonic term proportional to  $X_t^4$  should be transformed using the phase transformation $X_t= -\sqrt{(1+s)/2}\mathcal{X}_t+\sqrt{(1-s)/2}\mathcal{X}_r$, and this can make many types of terms possible, e.g. $C_{m} \mathcal{X}_r^m \mathcal{X}_t^{4-m}$ with  coupling strengths $C_{m}(s)$ and $m=0,1,2,3,4$.  In the  original eigenbasis $|n_j \rangle$, with $j=t,r$,  the ladder operators $\hat{a}_j=\sum_{n_j} \sqrt{n_j+1} \ket{n_j}\bra{n_j+1}$  can help to rewrite the charge operator $\hat{Q}_j= \sqrt{\hbar /2 \omega_j} (\oad_j + \oa_j) $ and the phase operator $\hat{X}_j=i\sqrt{\hbar \omega_j/2 } (\oad_j - \oa_j)$ \cite{Clerk}. Similarly in the normal mode basis the ladder operators $\hat{\alpha}_k$ determine the new charge and phase operators: $\hat{\mathcal{Q}}_j$ and $\hat{\mathcal{X}}_j$. These two basis can be transformed into one another using the following Bogoliubov-Velatin transformation: $\oad_t-\oa_t=U_{tt}(\oald_t - \oal_t) +U_{tr}(\oald_r - \oal_r)$, with $U_{tt}=-[{ (1+s){\bar{\omega}_t}/{2}{\omega_t}}]^{\frac{1}{2}}$ and $U_{tr}= [{ {(1-s)} {\bar{\omega}_r}/2{{\omega}_t}}]^{\frac{1}{2}}$.

The anharmonic quantum hamiltonian from Eq. (\ref{eq. Hexm1}) can be written  as $H_{\textup{anhar.}}=-(\delta/12) (\oad_t-\oa_t)^4$, with  $\delta\equiv  E_C$ being the anharmonicity coefficient.  In the new basis, this hamiltonian is transformed to $-\frac{\delta}{12} [U_{tt}(\oald_t - \oal_t) +U_{tr}(\oald_r - \oal_r)]^4$, defining self-Kerr coefficient \cite{Boissonneault} of the transmon ${\chi}_t=\delta U_{tt}^4 $ and that of the resonator ${\chi}_r=\delta U_{tr}^4 $.  Note that the anharmonic hamiltonian is not diagonal in the normal mode basis, however  we can simplify it  by ignoring irrelevant terms to first order and applying secular approximation. This reformulates total hamiltonian to  
\begin{eqnarray} \nonumber \label{eq. H_1t_1r}
H&=&\sum_{i=t,r}\bar{\omega}_{i} \oal_i^{\dagger} \oal_i-\frac{{\chi}_{i}}{2}   \left[ \left( \oal_i^{\dagger} \oal_i \right)^{2} + \oal_i^{\dagger} \oal_i +\frac{1}{2}\right]  \\ &&   - 2 {\chi}_{rt} \left( \oal_t^{\dagger} \oal_t  +\frac{1}{2}\right) \left( \oal_r^{\dagger}\oal_r  +\frac{1}{2}\right), 
\end{eqnarray}

The transmon state in normal mode basis makes a shift proportional to ${\chi}_{rt}$, namely the cross-Kerr coefficient, in the resonator frequency.  It is simple to show that  ${\chi}_{rt} = \sqrt{{\chi}_{r}{\chi}_{t}}$ and therefore it linearly scales with the anharmonicity $\delta$. Defining `dressed frequency' $\tilde{\omega}_i$ to be the coefficient of $\hat{\alpha}_i^\dagger \hat{\alpha}_i$, after summing over all relevant terms we find the following closed form formula for the dressed frequencies: 
\begin{eqnarray}  
&& \tilde{\omega}_{t} =  K_-^2 -\frac{{\chi}_t}{2}-{\chi}_{rt}, \ \ \ \ \tilde{\omega}_{r}  =  K_+^2 -\frac{{\chi}_r}{2}-{\chi}_{rt}  \label{eq. freqs}
\\ &&  {\chi}_{t} =  \delta{\left(1+s\right)^{2}}K_{-}^4/{4\omega_t^{2}}, \ \  {\chi}_{r}  =  \delta {\left(1-s\right)^{2}}{K_{+}^{4}}/{4 \omega_t^2}\ \ \  \ \   \label{eq. kerr}
\end{eqnarray}
which indicate $E_{n_t n_r} = \sum_{i}\tilde{\omega}_i n_i-\chi_i^2 n_i^2/2-2 \chi_{rt} n_t n_r$. The validity of these formulas are much wider than dispersive regime, in fact they are valid for \emph{arbitrary} coupling strength and frequency detuning.  

Let us compare our results with other models.  In general two sets of analytical results are known for the circuit: i) within the dispersive regime, which defines the validity of perturbation theory, or ii) at resonant frequencies in the Jaynes-Cummings model.  In dispersive regime the detuning  frequency is much stronger than the coupling strength, i.e. $g/\Delta \ll 1$, and within this regime  Eqs.  (\ref{eq. freqs}) and (\ref{eq. kerr}) are expanded in polynomials of $g/\Delta$. This will result in the following dressed frequencies:  $\tilde{\omega}_{t} \approx \omega_{t} - 2 \omega_r  {g}/{\Delta}\Sigma$, $ \tilde{\omega}_r  \approx \omega_{r}+ 2 {g^2}\omega_t/{\Delta}\Sigma  $ and the self-Kerr nonlinearities ${\chi}_{t}  =  \delta [1- 4  {g}^2 \omega_r (\omega_r^2+\omega_t^2) /\omega_t \Sigma^2 {\Delta}^{2}]$ and   ${\chi}_{r} =   16 \delta \left({g}/{\Delta}\right)^{4} \omega_r^4 / \Sigma^4 $.  These expression are in agreement with  the non-Rotating Wave Approximation (non-RWA)  results recently reported in \cite{Gely} using perturbation theory.  In the RWA regime $g\ll \Delta \ll \Sigma$, therefore we can  simplify these expressions by using \cite{Gely}: $\omega_r/\Sigma \approx \omega_t /\Sigma \approx 1/2$. The RWA dressed frequencies are  $\tilde{\omega}_{t}^{\textup{RWA}} \approx \omega_{t}-{\delta}/{2} -   {g^2}/{\Delta} - \delta g^2/ \Delta^2 $,  $\tilde{\omega}_r^{\textup{RWA}}  \approx \omega_{r}+  {g^2}/{\Delta}  - \delta g^2/\Delta^2 $, ${\chi}_{t}^{\textup{RWA}}  \approx  \delta [1- 2  ({g}/{\Delta})^{2} ]$,    ${\chi}_{r}^{\textup{RWA}} \approx   \delta \left({g}/{\Delta}\right)^{4} $.  These results are in agreement with the original perturbative Lamb and AC-Stark shifts  reported by J. Koch, et.al. in Ref. \cite{Koch07} and experimentally observed  \cite{Fragner}.

\begin{figure}[t]
\begin{center}
 \includegraphics[width=0.92\linewidth]{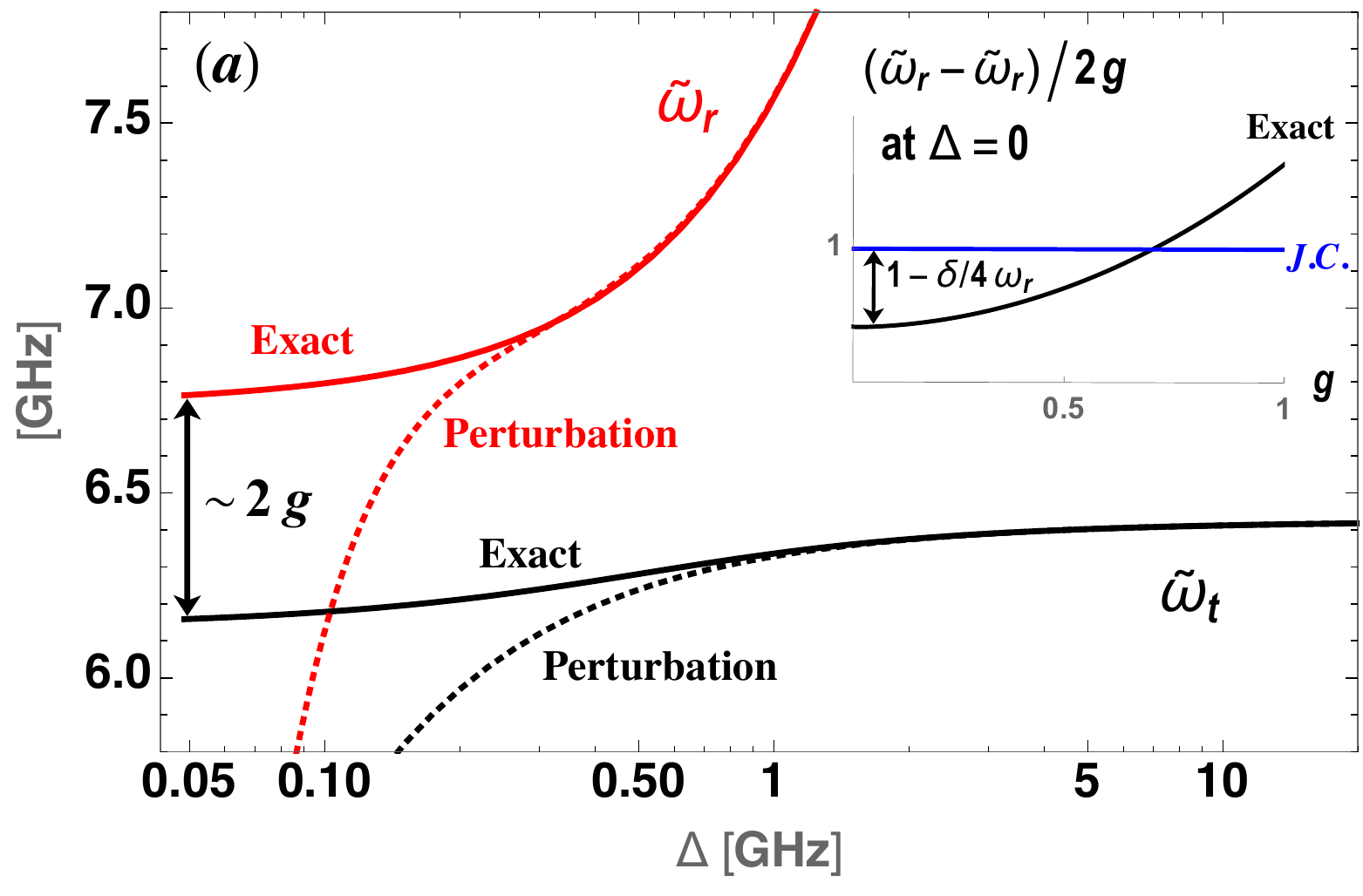} \\  
  \includegraphics[width=0.92\linewidth]{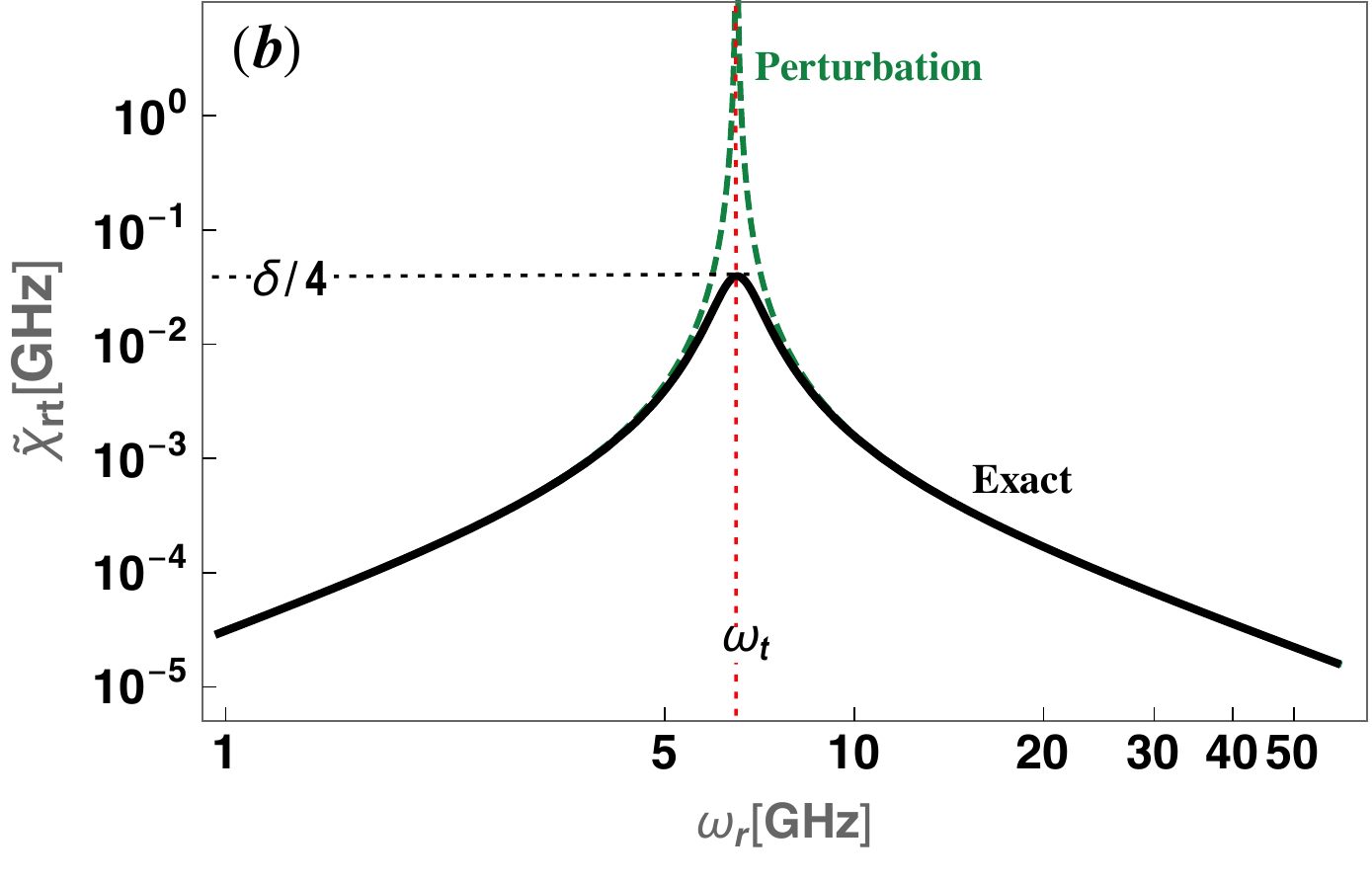} 
\caption{Exact (solid) and perturbative (dashed) results for (a) dressed frequencies  (b)  cross Kerr in circuit with $\omega_t=6.5$GHz, $\omega_r=\Delta+\omega_t$, $g=0.3$GHz, and $\delta=0.15$GHz. (Inset) Rescaled dressed frequency detuning  at resonant bare frequency in the Jayns-Cumming and the exact  models.}
\label{fig 1}
\end{center}
\end{figure}

Fig(\ref{fig 1}a) shows the transmon and the resonator dressed frequencies at different detuning frequencies $\Delta$ and a fixed coupling strength $g$. For the choice of circuit parameters, there is a negligible mismatch between RWA and non-RWA perturbative results, therefore,  in the logarithmic scales the lines labelled by Perturbation can be plotted using both formulations. In large detuning $\Delta$ the exact dressed frequencies of Eq. (\ref{eq. freqs}), in solid lines, are in good agreement with the perturbative (dotted) results.   However, as $g/\Delta$ increases almost above $\sim$ 1/3 the difference between exact and perturbative results starts to appear.  Another regime of interest is the special solution of the resonant point, where bare frequency of the transmon and the resonator are the same. Perturbation theory at this point diverges, however, the Jaynes-Cumming model predicts that due to atom-photon coupling a $2g$ frequency gap between the two dressed frequencies is produced \cite{{Haroche},{Blais04}}.  Our exact formalism in Eq. (\ref{eq. freqs}) not only confirms this result but also provides a modification in it due to the presence of finite anharmonicity in transmons, which makes the dressed frequency detuning $2g(1 - \delta /4\omega_r)$.  Fig. (\ref{fig 1}b inset) shows this gap rescaled by $2g$ at different coupling strengths $g$, which is unity for infinite anharmonicity (labelled J.C.); and is nonlinear for finite anharmonicity.  Fig. (\ref{fig 1}b) shows cross-Kerr coefficient --defined below Eq. (\ref{eq. H_1t_1r})-- in solid line and compares it with the perturbative results in dashed lines. At the resonant point as expected the perturbation theory diverges, however, in contrast, the exact solution reveals the finite value of $\delta/4+o(g^2)$ for any choice of bare frequencies.

\section{General method: N-atoms coupled to a resonator}

\emph{$N$ transmons coupled to $M$ resonator. --} Black box quantization in the original form has been proposed as an experimental method to get theoretical feedback on fitting parameters. Here we study a purely theoretical approach to generalize it to $N$ modes and $M$ transmons. This will help not only to scale up quantum circuits, but also it helps to study nonperturbative solutions. 

Scaling up entanglement is one of the purely quantum phenomenon that is most crucial for quantum computing. Such phenomenon can take place in large scale quantum circuits with $N$ transmons interacting with $M$ resonators.  A total of $N+M$ pairs of  canonical variables can be defined:  the charge vector $\mathbf{Q}=\left(Q_{1},\cdots, Q_{N+M}\right)^{T}$ and the  phase vector $\mathbf{X} = \left(X_{1}, \cdots, X_{N+M} \right)^{T}$. The circuit hamiltonian can be divided into a harmonic sector and a weakly anharmonic sector. The harmonic hamiltonian is $H_{\textup{har.}}=\frac{1}{2}\sum_{i=1}^{N+M} \omega_i^2 Q_{i}^{2}+\frac{1}{2} X_{i}^{2}+\sum_{i=1}^{N} \sum_{j=N+1}^{N+M} g_{ij}\sqrt{4\omega_{i}\omega_{j}}Q_{i}Q_{j}$. Using a generalization of $\mathbf{M}$ matrix in Eq. (\ref{eq:M}) this hamiltonian is simplified to $H_{\textup{har.}}=\frac{1}{2}\mathbf{Q}^\textup{T} \mathbf{M Q}+\frac{1}{2}\mathbf{X}^\textup{T}\mathbf{  X}$  with the matrix $\mathbf{M}$ being nonzero only at $M_{RR}=\omega_r^2$,  $M_{TT}=\omega_a^2$, $M_{TR}=M_{RT}=g_{t}\sqrt{4 \omega_t \omega_r}$;  Sub-indices $_T$ labels  transmons $\{ 1,2, \cdots,N\}$ and $_R$ the resonators $\{N+1,\cdots,M\}$. Consider that the following unitary transformations charges ${Q}_{i}=\sum_{j}S_{ij}{\mathcal{Q}}_{j}$ and phases ${X}_{i}=\sum_{j}T_{ij}{\mathcal{X}}_{j}$ take them to a normal mode basis. As  discussed in Appendix \ref{sec.Unitary_SM}, these unitary transformations are identical, i.e. $T_{ij}=S_{ij}$. They transform the harmonic hamiltonian to $\frac{1}{2}\sum_{i}\bar{\omega}_i\mathcal{Q}_{i}^{2}+\mathcal{X}_{i}^{2}$.  Detailed analysis show that $S$ is the matrix of normalized eigenvectors of  $\mathbf{M}$-matrix. This evaluates dressed frequencies in the absence of anharmonicity, which in this paper we determine them exactly in closed form formula for circuits with one and two transmons and one resonator, however for larger circuits the M-matrix  can be evaluated numerically and this determines all exact dressed frequencies.  

Once $S$ is found charges and phases can be promoted to operators and rewritten in terms of ladder operator  $\hat{a}$ in the original basis and $\hat{\alpha}$ in the new basis. They transform to one another as follows:
\begin{eqnarray}
\hat{a}^\dagger_{i}-\hat{a}_i & = & \sum_{j=1}^{N+M}  U_{ij}\left( \hat{\alpha}^\dagger_{j}- \hat{\alpha}_j \right), \ \ \ U_{ij}\equiv \sqrt{\frac{\bar{\omega}_j}{\omega_i}} S_{ij}  \label{eq. Bog}  
\end{eqnarray}

The anharmonic hamiltonian $\sum_{i=1}^{N}(\delta_i/12)(\hat{a}_i-\hat{a}_i^\dagger)^4$ can be similarly taken to the normal mode basis --see appendix \ref{sec.Bogoliuboc_SM} for details. The smallness of anharmonicity in transmons allows the nonlinear physical parameters to be evaluated in leading order.

\section{Two transmons coupled to a resonator}

\emph{Two transmons sharing a bus resonator --} This is an important circuit for two-qubit gate calibration \cite{{Blais07},{Majer07}}.   Let us denote $\omega_i$ with $i=1,2,3$ for the two transmons, and the resonator, respectively --alternatively we sometimes use $r$ (instead of $3$) to emphasize the resonator.   The coupling strengths between the transmons and the resonator are  $g_{1}, g_{2}$. The $\mathbf{M}$-matrix is \begin{equation}
\mathbf{M}=\left[\begin{array}{ccc}
\omega_{1}^{2} & 0 & g_{1}\sqrt{4\omega_{1}\omega_{3}}\\
0 & \omega_{2}^{2} & g_{2}\sqrt{4\omega_{2}\omega_{3}}\\
g_{1}\sqrt{4\omega_{1}\omega_{3}} & g_{2}\sqrt{4\omega_{2}\omega_{3}} & \omega_{3}^{2}
\end{array}\right]\label{eq:M3dar3}
\end{equation}

The M-matrix can be taken to a normal mode basis within a wide domain of parameters that includes the superconducting circuits of interest for  quantum computation --see below Eq. (\ref{eq:eignfreq}). The cubic equation $\lambda^{3}+b\lambda^{2}+c\lambda+d=0$ determines the eigenvalues $\lambda$ of Eq. (\ref{eq:M3dar3})  with $b\equiv -\sum_{i=1,2,3}\omega_{i}^{2}$,
$c\equiv  \omega_{1}^{2}\omega_{2}^{2}+\omega_{1}^{2}\omega_{3}^{2}+\omega_{2}^{2}\omega_{3}^{2}-\sum_{i=1,2} 4g_{i}^{2}\omega_{i}\omega_{3}$,
and $d\equiv  4g_{2}^{2}\omega_{1}^{2}\omega_{2}\omega_{3}+4g_{1}^{2}\omega_{1}\omega_{2}^{2}\omega_{3}-\omega_{1}^{2}\omega_{2}^{2}\omega_{3}^{2}$. Note that the eigenvalues $\lambda_k$ determine the circuit dressed frequencies, i.e. $\bar{\omega}_{k}\equiv \sqrt{\lambda_k}$, with  $k=1,2,3$. By defining $\theta\equiv \lambda+b/3$ the quadratic term is eliminated, i.e. $\theta^{3}-f\theta+h=0$, $f\equiv  b^{2}/3-c$ and $h\equiv  \left(2b^{3}-9bc+27d\right)/27$. We solve this equation using trigonometric trials functions and find  
\begin{equation}
\bar{\omega}_{k}^2 =  {2\sqrt{\frac{f}{3}}\cos \frac{ \textup{cos}^{-1}\left(-\frac{h}{2}\left(\frac{3}{f}\right)^{\frac{3}{2}}\right)-2\pi(k-1)}{3}-\frac{b}{3}}  \label{eq:eignfreq}
\end{equation}
A relabelling of indices might be needed to identify corresponding frequencies. Should the inverse cosine be always between -1 and 1, ${h^{2}}/{4}-{f^{3}}/{27}<0$ must be satisfied for real-valued solutions. In circuits suitable for quantum computation, however, since coupling strengths are much smaller than individual frequencies this condition is trivially satisfied  Appendix \ref{sec.limits_SM}.

\begin{figure}[t]
\begin{center}
 \includegraphics[width=0.77\linewidth]{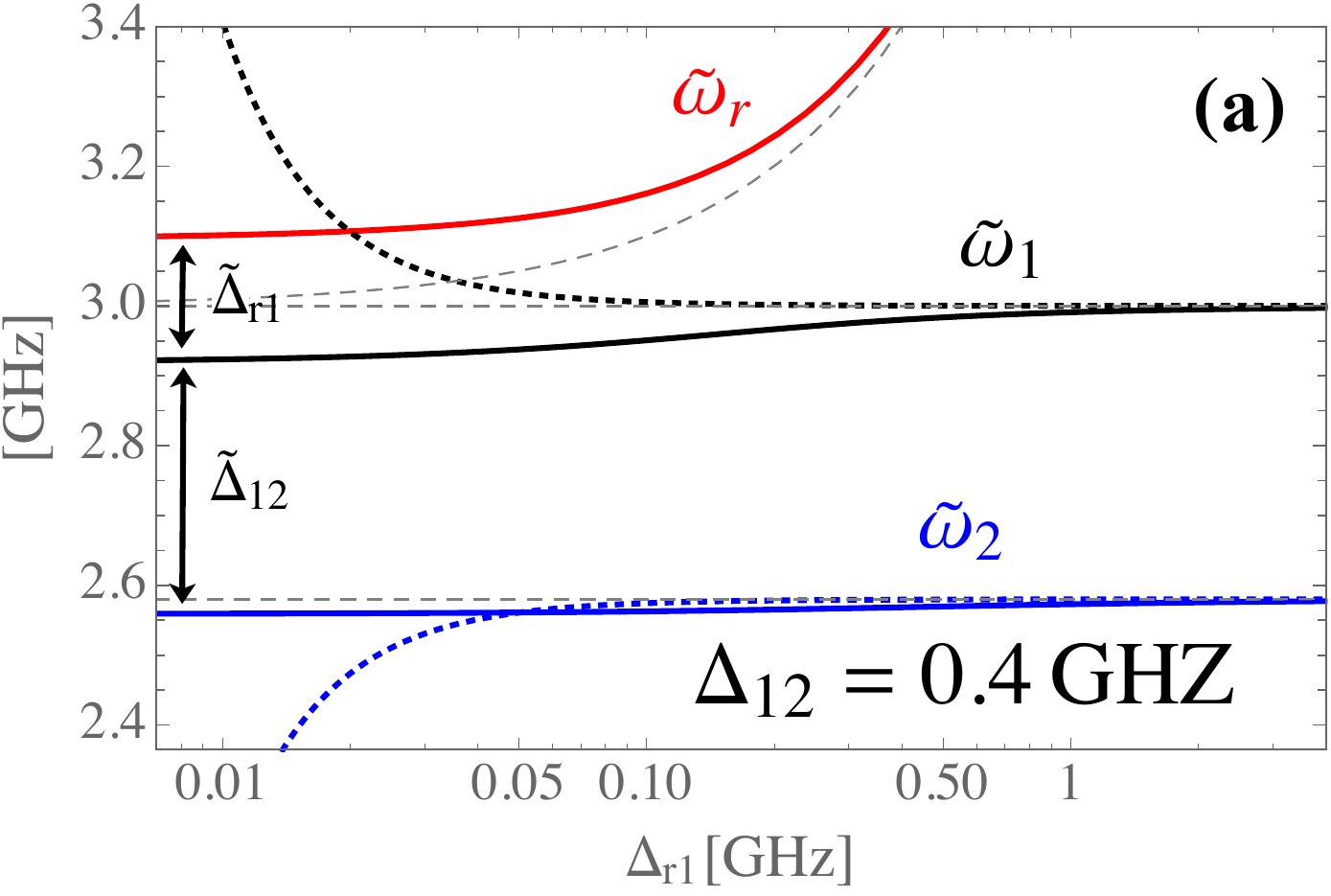} \\ 
  \includegraphics[width=0.77\linewidth]{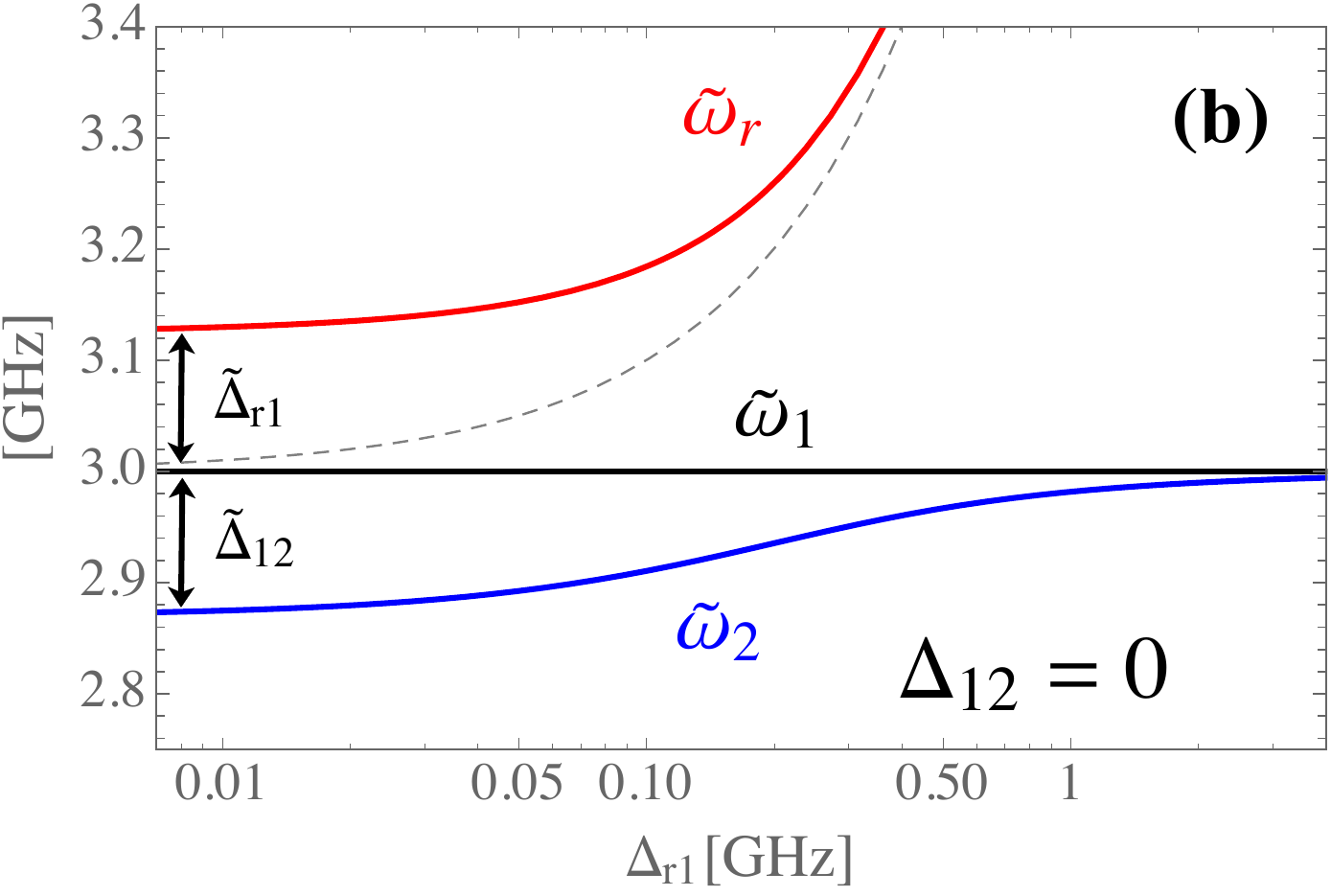} 
\caption{Perturbative (dotted) and exact (solid)  dressed frequencies in circuit with  bare frequencies (dashed) of transmons $\omega_{1}=3$GHz, $\omega_2=\alpha \omega_1$, and the resonator $\omega_r=\omega_1+\Delta_{r1}$, couplings $g\equiv g_{1,2}$(=0.1GHz) and  $\delta\equiv \delta_{1,2}$(=0.1GHz). (a) $\alpha=0.86$, and (b) three-body resonance $\alpha=1$. }
\label{fig 3}
\end{center}
\end{figure}

The anharmonic hamiltonian transformed into the normal mode basis in the leading order can be written as 
\begin{eqnarray} \nonumber
&&H=\sum_{i=1,2,3} \tilde{\omega}_i \hat{\alpha}_i^\dagger \hat{\alpha}_i       \\ \nonumber &&
  - \sum_{i=1,2,3}\Bigg\{ \frac{\chi_{i}}{2}\left(\hat{\alpha}_{i}^{\dagger}\hat{\alpha}_{i}\right)^{2} +  2 \sum_{k > i} \chi_{ik}\left(\hat{\alpha}_{i}^{\dagger}\hat{\alpha}_{i}\right)\left(\hat{\alpha}_{k}^{\dagger}\hat{\alpha}_{k}\right)  \Bigg.  \\ \nonumber &&
 \ \ \ +   \sum_{k > i} \left( \mathcal{J}_{ik}+ \sum_{l\neq i,k} \mathcal{L}_{ikl} \hat{\alpha}_{l}^{\dagger}\hat{\alpha}_{l}  \right) \left(\hat{\alpha_{i}}\hat{\alpha}_{k}^{\dagger}+\hat{\alpha}_{i}^{\dagger}\hat{\alpha}_{k}\right)
   \\  && \Bigg.  \ \ \ + \sum_{k\neq i} \mathcal{K}_{ik}  \left[\left(\hat{\alpha}_{i}^{\dagger}\hat{\alpha}_{i}\right)\hat{\alpha}_{i}\hat{\alpha}_{k}^{\dagger}+\hat{\alpha}_{i}^{\dagger}\hat{\alpha}_{k}\left(\hat{\alpha}_{i}^{\dagger}\hat{\alpha}_{i}\right)\right] \Bigg\}
  \label{eq. 2trans}
 \end{eqnarray}
with self-Kerr $\chi_{i}=\sum_{j=1,2} \delta_j U_{ji}^{4}$ and  cross-Kerr $\chi_{ik}=\sum_{j=1,2} \delta_j U_{ji}^{2} U_{jk}^{2}$ and $U_{ij}$ being define in Eq. (\ref{eq. Bog}) ---see Appendix \ref{sec. trans 2transmons_SM} for details.   One can  evaluate  the  Kerr cofactors and see that  in general there is no simple relation between cross Kerr  and self-Kerr coefficients.

In Eq. (\ref{eq. 2trans}) the   $\mathcal{J}$ coupling indicates a direct interaction between two oscillators. The  $\mathcal{K}$ and $ \mathcal{L}$ terms are multiplied by $\hat{\alpha}^\dagger_i \hat{\alpha}_i$, therefore they are effectively $ n_l \mathcal{L}_{ikl} $ and $n_i \mathcal{K}_{ik} $ with $n$ being  integer quantum numbers. These couplings linearly depend on anharmonicity $\delta$ and are stronger in higher excited states. Detailed analysis ---see Appendix \ref{sec_Additional} and \ref{sec. SW_SM}--- show that after block-diagonalization these three interactions appear in the effective hamiltonian of the two-qubit circuit only in the higher orders $\delta^2$, thus they are negligible in the leading order.
 
The dressed frequency of   transmons and  resonator are 
\begin{equation}
\tilde{\omega}_{i} = \bar{\omega}_{i} -\frac{{\chi}_i}{2}-\sum_{j  (\neq i)}{\chi}_{ij}. \label{eq. dress_w_2tr}
\end{equation}
and the energy levels are $E_{n_1 n_2 n_3} = \sum_{i=1}^3 n_i \tilde{\omega}_{i}-\chi_i n_i^2/2-2\sum_{k> i}\chi_{ik} n_i n_k $.  Fig. (\ref{fig 3}a) shows all dressed frequencies given the bare values $\omega_1=3$GHz, $\omega_2=\alpha \omega_1$, and $\omega_r=\omega_1+\Delta_{r1} $.  We obtain perturbative results (dotted)  using the formalism explained in Ref. \cite{{Magesan},{Gambetta13}} ---more explicitly Eqs. (4.3-4.5) of the first reference. For the fixed coupling strength $g$ and large $\Delta_{r1}\gg  g$  the results of perturbation theory and Eq. (\ref{eq. dress_w_2tr}) are in good agreement. Far from the dispersive regime, however, the two become much deviated. For example in a  circuit with the frequency of transmon 1 and resonator in resonance perturbation theory diverges, however Eq. (\ref{eq. dress_w_2tr}) predicts a finite dressed frequency gap as shown in Fig. (\ref{fig 3}a). In the case of $\alpha<1$ and $g/\omega_1  < (1-\alpha^2)/3\sqrt{6}$, a series expnasion of Eq. (\ref{eq. dress_w_2tr})  in terms of $g$ shows that in absence of anharmonicity $\tilde{\Delta}_{r1}\approx 2g$ and $\tilde{\Delta}_{12} \approx \Delta_{12}-g + [(1+\alpha)^{-1}+(1+\alpha)/2]g^2/\Delta_{12}+o(g^3)$.  Fig. (\ref{fig 3}b) shows the special case of resonant transmons with $\omega_1=\omega_2\equiv \omega$. Our exact evaluation indicates that all dressed frequencies become off resonant at the values $\omega$, $\sqrt{({\omega^2+\omega_r^2 \pm \Delta \Sigma r^{-1}})/2}$ with $r^{-2}\equiv 1+32 g^2 \omega \omega_r/\Delta_{r1}^2 \Sigma_{r1}^2$. The special case of maximal resonance, i.e. $\omega_r=\omega$, the dressed frequencies will be found  $\tilde{\Delta}_{r1} \approx \tilde{\Delta}_{12}\approx \sqrt{2} g$ in absence of anharmonicity, see Appendix \ref{sec. resonance_SM} for details. These are all previously unknown results that can be used for instance to identify bad samples in circuit fabrications. 

Before conclusion let us comment on evaluating the effective impedances introduced by the black box quantization \cite{Nigg12}. They are assumed to be unknown parameters and can be evaluated in iterative feedbacks between theory and experiment. Our formalism, however,   reveals a number of constraints that link between the effective impedances that makes the simpler to be theoretically estimated. For the simple example of a transmon coupled to a resonator, the effective impedances can be found analytically:  $Z_t^{\textup{eff}}=\tilde{\omega}_t {(1+s)Z_{t}}/ {2\omega_t}$ and  $Z_r^{\textup{eff}}=\tilde{\omega}_r {(1-s)Z_{t}}/{2\omega_r}$. Their ratio in the dispersive regime is $Z_r^{\textup{eff}}/Z_t^{\textup{eff}} \sim (g/\Delta)^{2}$, which indicates the characteristic impedance associated to the transmon exceeds that of the resonator.

\section{Discussion}

We presented a rigorous method to exactly obtain effective qubit parameters from the hamiltonian of superconducting circuits consisting of resonators and JJs at arbitrary coupling strengths and frequency detunings. Using this formalism we exhibited single transmon and two transmons outside of the dispersive regime in closed-form formulas.  For complicated circuits finding analytical expressions may not be easy,  however,  our formalism can determine qubit parameters numerically much easier and more accurate compared to perturbation theory in charge basis, because the M-matrix (defined in the text) linearly  scales with the number of qubits and resonators and all we need is to find its eigenvectors. This simplicity and accuracy will play an essential role for scaling up superconducting circuits as it allows to explore the possibilities of new domains of parameters for elevated fidelities.

\begin{acknowledgments}
We thank David DiVincenzo for many useful discussions. Support from Intelligence Advanced Research Projects Activity (IARPA) under contract W911NF-16-0114 is gratefully acknowledged.
\end{acknowledgments}

\section*{Author contributions statement}

M.A. conducted  theoretical analyses. 

\section*{Additional information}

\textbf{Competing interests} 
The authors declare no competing interests.

\appendix


\section{Unitary transformation of canonical variables}
\label{sec.Unitary_SM}
Consider two $N$ dimensional vectors of canonical variables $\mathbf{q}= (q_1, q_2, \cdots, q_N )$ and $\mathbf{p}=(p_1, p_2, \cdots, p_N )$. These variables satisfy the Poisson bracket relation $\{ q_i,p_j\}=\delta_{ij}$ with $i,j=1,2,\cdots,N$ and the definition of $\{f,g\}=\sum_{i=1}^N (\partial f/\partial q_i) (\partial g/\partial p_i) - (\partial f/\partial p_i) (\partial g/\partial q_i)$. 

Let us consider the following unitary transformations takes place on these variables: $Q_i=\sum_{j=1}^N S_{ij} q_j$ and $P_i=\sum_{j=1}^N T_{ij} p_j$.  In order to have the two new variables $\mathbf{Q}$ and $\mathbf{P}$ to be canonical variables they must satisfy similar Poisson bracket relation as those of old variables:  $\{ Q_i,P_j\}=\delta_{ij}$. This indicates that $\{Q_i,P_j\}=\sum_{k=1}^N (\partial Q_i/\partial q_k) (\partial P_j/\partial p_k) - (\partial Q_i/\partial p_k) (\partial P_j/\partial q_k)$. One can easily simplify these relations into: $\sum_{k=1}^N S_{ik} T_{jk} =\delta_{ij}$. Because of the unitarity of the transformation matrices $S$ and $T$ one can see that $\sum_{k=1}^N S_{ik} S_{kj}^\dagger =\delta_{ij}$. For real matrices we have $S^\dagger_{kj}=S_{jk}$, thus $\mathbf{T}=\mathbf{S}$.

%
%


\section{Constraints within exact formula for 2 transmon circuit}
\label{sec.limits_SM}

Another  condition that can be concluded from  Eq. (8) of the main article is the following: 
\begin{equation}
{2\sqrt{\frac{f}{3}}\cos \frac{ \textup{cos}^{-1}\left(-\frac{h}{2}\left(\frac{3}{f}\right)^{\frac{3}{2}}\right)-2\pi(k-1)}{3}- \frac{b}{3}}\geq 0  \label{eq:cond_SM}
\end{equation}

By definition we have always $b\leq 0$, therefore the condition can be checked in the cases where $\cos$ function is negative, therefore we need to check   the following condition: $-2\sqrt{f/3}+ |b|/3\geq 0 $, which can be further simplified to ${b^2}/{3}>c$.  Substituting the definitions will introduce the following condition to hold:
\[ \omega_1^4+\omega_2^4 + \omega_3^4 \geq \omega_1^2 \omega_2^2 + \omega_1^2 \omega_3^2 + \omega_2^2 \omega_3^2 - 4 g_1^2 \omega_1 \omega_2 - 4 g_2^2 \omega_2 \omega_3 \]

We take first three terms from right side to the left, then simplify left side to arrive at the following condition: 
\[ (\omega_1^2 - \omega_2^2)^2 + (\omega_2^2 - \omega_3^2)^2 + (\omega_3^2 - \omega_1^2)^2\geq  - 4 g_1^2 \omega_1 \omega_2 - 4 g_2^2 \omega_2 \omega_3 \]
which trivially holds valid without imposing any limitations on parameters.


\section{Unitary transformation for 2 transmons coupled to resonator}
\label{sec. trans 2transmons_SM}
The unitary transformation to diagonal basis in the harmonic sector is carried out by the matrix of normalized eigenstates with columns being eigenvectors, which is 
\begin{equation}
S=\left[\begin{array}{ccc}
\frac{V_1 \gamma_{12}}{N_1} & \frac{V_1 \gamma_{22}}{N_2} & \frac{V_1 \gamma_{32}}{N_3}\\
\frac{V_2 \gamma_{11}}{N_1} & \frac{V_2 \gamma_{21}}{N_2} & \frac{V_2 \gamma_{31}}{N_3}\\
\frac{\gamma_{11} \gamma_{12}}{N_1} & \frac{\gamma_{21} \gamma_{22}}{N_2} & \frac{\gamma_{31} \gamma_{32}}{N_3}
\end{array}\right]\label{eq:Msupp}\end{equation}
with $V_i\equiv g_i\sqrt{4 \omega_r \omega_i}$, $\gamma_{ij}=\bar{\omega}_i^2-\omega_j^2$, and $N_i=\sqrt{V_2^2 \gamma_{i1}^2+V_1^2 \gamma_{i2}^2+\gamma_{i1}^2 \gamma_{i2}^2}$.


\section{Additional interaction terms}
\label{sec_Additional}
In the circuit made of two transmons coupled to a shared resonator, the anharmonic part of Hamiltonian can be simplified to Eq. (9) in the main article. Below are detailed interaction couplings in terms of bare parameters:
\begin{eqnarray} \nonumber
&& \mathcal{J}_{ik}  = \sum_{j=1,2} \delta_j \left[\frac{1}{3} U_{ji}^{3} U_{jk} + U_{jk} ^{3} U_{ji} + \frac{2}{3} \frac{(U_{ji}U_{jk}U_{j3})^{2} }{U_{j1}U_{j2} } \right], \\ \nonumber
&& \mathcal{K}_{ik} = \sum_{j=1,2} \delta_j U_{ji}^{3}U_{jk} , \ \ \  \mathcal{S}_{ikl} = \frac{4}{3}\sum_{j=1,2} \delta_j \frac{(U_{ji}U_{jk}U_{j3})^{2}}{U_{ji}U_{jk}}, \\
\label{eq. coup_SM}
\end{eqnarray}



\section{Block diagonalization}
\label{sec. SW_SM}
Let us consider the the Hamiltonians of two harmonic oscillators (labeled as 1, 2) coupled to a resonator (labeled as 3):
\begin{eqnarray*} \nonumber
&& H=H_0+\epsilon H_{int},\ \ \ \  H_0\equiv \sum_{i=1,2,3}  \omega_i \hat{\alpha}_i^\dagger \hat{\alpha}_i, \\    \nonumber 
&&   H_{int}\equiv  \sum_{k=1,2} g_k  \left(\hat{\alpha}_{3}\hat{\alpha}_{k}^{\dagger}+\hat{\alpha}_{3}^{\dagger}\hat{\alpha}_{k}\right) \\
  \label{eq. H_SM}
 \end{eqnarray*}

The unperturbed part $H_0$ in the eigenbasis of itself id diagonal, however $H_{int}$ is not. In general we may not be able to find a tranformation to fully diagonal matrix, but instead we can separate out a subset of states from the rest of the states.  The  Schrieffer-Wolff transformation is one way to  block diagonalize the interacting Hamiltonian into  low energy and  high energy sectors.  This usually takes place by transforming the Hamiltonian by the anti-hermitian operator $\exp{S}$ in the following way:  $H_{BD}=\exp{(-S)} H \exp{S}$, which can be expanded into $H_{BD}=\sum_{n=0} [H,S]_{n}/n!$ with $[H,S]_{n+1}=[[H,S]_{n},S]$ and $[H,S]_{0}=H$.  One can in principle assume a geometric series expansion of the transformation matrix:   $S=\sum_{i=0} (\epsilon )^i S_i$; however  given that the zeroth or order of $H_{BD}$ is $H_{{BD}_0}=[H_0,S_0]=H_0$ therefore $S_0$ must be diagonal too which is in fact inconsistent with the definition of $S$ to be anti-hermitian and block-off-diagonal, therefore always $S_0=0$.  In the first order the Hamiltonian is already given by $H_{int}$ which can be made of block-diagonal (bd) and block-off-diagonal (bod) matrices $H_{int}=H_{int}^{\textup{bd}}+H_{int}^{\textup{bod}}$ . Therefore $H_{BD_1}=[H_0,S_1]=-H_{int}^{\textup{bod}}$. In the second order: $H_{BD_2}=[H_0,S_2]+[H_{int},S_1]+(1/2)[[H_0,S_1],S_1]$, and so on.  Putting all together one can find the effective Hamiltonian up to the second order $H_{BD}=H_0+H_{int}^{\textup{bd}}+(1/2)[H_{int}^{\textup{bod}},S_1]$. 

Using the relations above for the Hamiltonian of Eq. (\ref{eq. H_SM}) in which the interaction is block-off-diagonal one can use the following ansatz 

\begin{equation}
S_1=- \sum_{k=1,2} g_{k}\left(\hat{\alpha_{3}}\hat{\alpha'}_{k}^{\dagger}-\hat{\alpha}_{3}^{\dagger}\hat{\alpha'}_{k}\right).
\end{equation} 
with $\hat{\alpha'}_{k}\equiv \sum_{n=0}^\infty \sqrt{n+1}(\omega_3-\omega_k)^{-1} |n\rangle \langle n+1|$ being the modified ladder operator for $k$-th transmon, given that the normal ladder operator for the same transmon is $\hat{\alpha}_{k}\equiv \sum_{n=0}^\infty \sqrt{n+1} |n\rangle \langle n+1|$ . 

One can explicitly determine the effective Hamiltonian up to the second order of perturbation theory  becomes
\begin{equation}
H_{BD}=H_0-\sum_{i,j=1,2; (i\neq j)} \frac{g_i g_j}{2} \left(\hat{\alpha_{i}}\hat{\alpha'}_{j}^{\dagger}+\hat{\alpha}_{i}^{\dagger}\hat{\alpha'}_{j}\right)
\label{eq.HBD_SM}
\end{equation}


\section{Resonant transmons}
\label{sec. resonance_SM}

In a circuit with two transmons in resonance $\omega_1=\omega_2\equiv \omega$ and homogeneous coupling and anharmonicity  $g_1=g_2\equiv g$  and $\delta_1=\delta_2\equiv \delta$ the harmonic Hamiltonian is $H_{\textup{har.}}=\frac{1}{2} \omega^2 (Q_{1}^{2}+Q_2^2)+\frac{1}{2} \omega_r^2 Q_{r}^{2}+\frac{1}{2} (X_1^2+X_2^2+X_{r}^{2})+ g \sqrt{4\omega \omega_{r}}(Q_{1}+Q_2)Q_{3}$.  Defining the vectors $\mathbf{Q}=(Q_1, Q_2, Q_r)^T$ and $\mathbf{P}=(P_1, P_2, P_r)^T$, this Hamiltonian can be rewritten as $H_{\textup{har.}}=\frac{1}{2}\mathbf{Q}^\textup{T} \mathbf{M Q}+\frac{1}{2}\mathbf{X}^\textup{T}\mathbf{  X}$  with the matrix $\mathbf{M}$ being 
\begin{equation}
\mathbf{M}=\left[\begin{array}{ccc}
\omega^{2} & 0 & V\\
0 & \omega^{2} & V\\
V& V & \omega_{r}^{2}
\end{array}\right]\label{eq:M3dar3_SM}
\end{equation}
with $V\equiv g\sqrt{4\omega\omega_{r}}$. 
Because the off diagonal elements are identical,  it is easy to find the eigenvalues, which are
\[  \omega,\ \ \  \sqrt{\frac{\omega^2+\omega_r^2 \pm \sqrt{(\omega^2-\omega_r^2)^2+8V^2}}{2}} \]
At the exterm resonance with $\omega_r=\omega$ the eigenenergies will become
\[   \omega,\ \ \  \omega \sqrt{1 \pm  \frac{2\sqrt{2} g }{\omega}} \]

In the limit of small coupling $g\ll \omega$ this can be simplified to 
\[   \omega,\ \ \   \omega \pm  \sqrt{2} g  
\]
\


\section{Anaharmonicity}
\label{sec. anharmonicity_SM}

Consider the following Bogoliubov transformations for transmon ladder operator:
\begin{equation}
\hat{a}_{n}  =  \sum_{m}A_{nm}\hat{\alpha}_{m}+B_{nm}\hat{\alpha}_{m}^{\dagger}
\label{eq:bog}
\end{equation}
and using the relation between transmon charge number and phase and the ladder operator $\hat{a}_{n}=\sqrt{\frac{\omega_{n}}{2}}\hat{q}_{n}+i\sqrt{\frac{1}{2\omega_{n}}}\hat{p}_{n}$, and its conjugate as well as similar in the transformed basis $\hat{\alpha}_{n}=\sqrt{\frac{\tilde{\omega}_{n}}{2}}\hat{Q}_{n}+i\sqrt{\frac{1}{2\tilde{\omega}_{n}}}\hat{P}_{n}$, one can find
\begin{eqnarray*}
&& A_{nm}  =  \left(\sqrt{\frac{\omega_{n}}{8\tilde{\omega}_{m}}}+\sqrt{\frac{\tilde{\omega}_{m}}{8\omega_{n}}}\right)S_{nm},\\ 
& & B_{nm}  =  \left(\sqrt{\frac{\omega_{n}}{8\tilde{\omega}_{m}}}-\sqrt{\frac{\tilde{\omega}_{m}}{8\omega_{n}}}\right)S_{nm}
\end{eqnarray*}
in which $\tilde{\omega}$ is the frequency in the transformed basis. 


The anharmonicity in Hamiltonian will be $-\frac{\delta_{i}}{12}\left(a_{i}-a_{i}^{\dagger}\right)^{4}$. The operator part can be Bogoliubov transformed to the new basis, keeping terms with as many creations as annihilations, ignoring frequencies: 
\begin{widetext}
\begin{eqnarray*}
&& \left(a_{n}-a_{n}^{\dagger}\right)^{4}  = \\ && 6\sum_{m=1}^{3}\left(A_{nm}-B_{nm}\right)^{4}\left[\left(\hat{\alpha}_{m}^{\dagger}\hat{\alpha}_{m}\right)^{2}+\hat{\alpha}_{m}^{\dagger}\hat{\alpha}_{m}\right]\\
 &  & +6\sum_{m<k}\left(A_{nm}-B_{nm}\right)^{2}\left(A_{nk}-B_{nk}\right)^{2}\left[\hat{\alpha}_{m}^{2}\hat{\alpha}_{k}^{\dagger2}+\hat{\alpha}_{m}^{\dagger2}\hat{\alpha}_{k}^{2}+4\hat{\alpha}_{m}^{\dagger}\hat{\alpha}_{m}\hat{\alpha}_{k}^{\dagger}\hat{\alpha}_{k}+2\hat{\alpha}_{m}^{\dagger}\hat{\alpha}_{m}+2\hat{\alpha}_{k}^{\dagger}\hat{\alpha}_{k}\right]\\
 &  & +4\sum_{m\neq k}\left(A_{nm}-B_{nm}\right)^{3}\left(A_{nk}-B_{nk}\right)\left(\hat{\alpha}_{m}^{2}\hat{\alpha}_{m}^{\dagger}\hat{\alpha}_{k}^{\dagger}+\hat{\alpha}_{m}^{\dagger2}\hat{\alpha_{m}}\hat{\alpha}_{k}+2\hat{\alpha}_{m}^{\dagger}\hat{\alpha}_{m}\hat{\alpha_{m}}\hat{\alpha}_{k}^{\dagger}+2\hat{\alpha}_{m}^{\dagger}\hat{\alpha}_{m}\hat{\alpha}_{m}^{\dagger}\hat{\alpha}_{k}+\hat{\alpha_{m}}\hat{\alpha}_{k}^{\dagger}+\hat{\alpha}_{m}^{\dagger}\hat{\alpha}_{k}\right)\\
 &  & +8\sum_{m\neq k\neq l}\left(A_{nm}-B_{nm}\right)^{2}\left(A_{nk}-B_{nk}\right)\left(A_{nl}-B_{nl}\right)\left(\hat{\alpha}_{m}^{2}\hat{\alpha}_{l}^{\dagger}\hat{\alpha}_{k}^{\dagger}+\hat{\alpha}_{m}^{\dagger2}\hat{\alpha_{l}}\hat{\alpha}_{k}+2\hat{\alpha}_{m}^{\dagger}\hat{\alpha}_{m}\hat{\alpha_{l}}\hat{\alpha}_{k}^{\dagger}+2\hat{\alpha}_{m}^{\dagger}\hat{\alpha}_{m}\hat{\alpha}_{l}^{\dagger}\hat{\alpha}_{k}+\hat{\alpha_{l}}\hat{\alpha}_{k}^{\dagger}+\hat{\alpha}_{l}^{\dagger}\hat{\alpha}_{k}\right)
\end{eqnarray*}
\end{widetext}


\section{Bogoliubov transformation for Hamiltonian diagonalization}
\label{sec.Bogoliuboc_SM}

In this section we use quantum Hamiltonian of a transmon coupled to a resonator is $H  =  4E_{c}n-E_{J}\cos\phi+H_\textup{res}$. Separating the harmonic sector and the anharmonic sector, and using Bogoliubov transformation we diagonalize the interacting harmonic sector into a diagonal quantum harmonic Hamiltonian. We find all Bogoliubov transformation coefficients, which turns out to be similar to the results we took from semiclassical analysis.  

Given that charge number operator is proportional to ladder operators $n  \sim  2^{-\frac{1}{4}}\left(a+a^{\dagger}\right)$ and phase is the conjugate variable $\phi\sim2^{\frac{1}{4}}\left(a-a^{\dagger}\right)$, and the resonator Hamiltonian is $H_\textup{res}=\omega_{r}b^{\dagger}b$, the circuit Hamiltonian can be written as $H  =  \omega_{q}a^{\dagger}a-\frac{\delta}{12}\left(a-a^{\dagger}\right)^{4}+\omega_{r}b^{\dagger}b+g\left(a+a^{\dagger}\right)\left(b+b^{\dagger}\right)$ with harmonic part being $H_\text{har}  = \omega_{q}a^{\dagger}a+\omega_{r}b^{\dagger}b+g\left(a+a^{\dagger}\right)\left(b+b^{\dagger}\right)$.  

We would like to Bogoliubov-transform the Hamiltonian into a diagonal Hamiltonian $\mathcal{H} $: 
\[\mathcal{H}  =  \tilde{\omega}_{q}\alpha^{\dagger}\alpha+\tilde{\omega}_{r}\beta^{\dagger}\beta-\frac{1}{12}\left({\chi}_{q}^{\frac{1}{4}}\left(\alpha-\alpha^{\dagger}\right)+{\chi}_{r}^{\frac{1}{4}}\left(\beta-\beta^{\dagger}\right)\right)^{4}\]

We use a technique widely used in second quantized QFT, which is to
Bogoliubov-transofmation creation and annihilation operators
\[ \hat{a}  =  A\hat{\alpha}+B\hat{\beta}+C\alpha^{\dagger}+D\beta^{\dagger}, \ \ \ \ \hat{b}  =  E\hat{\alpha}+F\hat{\beta}+G\alpha^{\dagger}+H\beta^{\dagger}\]

Eight equations are needed to determines coefficients; four by enforcing
that transformed Hamiltonian preserves eigenvalues, which is equivalent
to equating $H_{ho}$ and $\mathcal{H}_{ho}$ and setting coefficients
of $\hat{\alpha}\hat{\alpha},\hat{\beta}\hat{\beta},\hat{\alpha}\hat{\beta}$
and $\hat{\alpha}\hat{\beta}^{\dagger}$to zero, respectively:
\begin{eqnarray}
 &  & \omega_{q}AC^{*}+\omega_{r}EG^{*}+g\left(A+C^{*}\right)\left(E+G^{*}\right)=0\label{eq:1-1}\\
 &  & \omega_{q}BD^{*}+\omega_{r}FH^{*}+g\left(B+D^{*}\right)\left(F+H^{*}\right)=0\label{eq:2-1}\\
 &  & \omega_{q}\left(BC^{*}+AD^{*}\right)+\omega_{r}\left(FG^{*}+EH^{*}\right)\label{eq:3-1}\\
 &  & \quad\quad+g\left[\left(A+C^{*}\right)\left(F+H^{*}\right)+\left(B+D^{*}\right)\left(E+G^{*}\right)\right]=0\nonumber \\
 &  & \omega_{q}\left(DC^{*}+AB^{*}\right)+\omega_{r}\left(HG^{*}+EF^{*}\right)\label{eq:4-1}\\
 &  & \quad\quad+g\left[\left(A+C^{*}\right)\left(F+H^{*}\right)+\left(B+D^{*}\right)\left(E+G^{*}\right)\right]=0\nonumber 
\end{eqnarray}
The other four are determined by enforcing commutation relations,
i.e. $[a,a^{\dagger}]=[b,b^{\dagger}]=1$ and $[a,b]=[a,b^{\dagger}]=0$,
respectively, given that $[\alpha,\alpha^{\dagger}]=[\beta,\beta^{\dagger}]=1$
and zero otherwise:
\begin{eqnarray}
|A|^{2}+|B|^{2}-|C|^{2}-|D|^{2} & = & 1,\label{eq:5-1}\\
|E|^{2}+|F|^{2}-|G|^{2}-|H|^{2} & = & 1,\label{eq:6-1}\\
AG+BH-CE-DF & = & 0,\label{eq:7-1}\\
AE^{*}+BF^{*}-CG^{*}-DH^{*} & = & 0.\label{eq:8-1}
\end{eqnarray}

For simplicity we assume coefficients are real-valued, but the equations are difficult to be analytically solved. A practical simplification can be achieved by defining new variables
\[ A_{\pm}\equiv A\pm C,\;B_{\pm}\equiv B\pm D,\;E_{\pm}\equiv E\pm G,\;F_{\pm}\equiv F\pm H \]
 which reformulates equations given above to the followings:
\begin{eqnarray*}
 &  & \omega_{q}\left(A_{+}^{2}-A_{-}^{2}\right)+\omega_{r}\left(E_{+}^{2}-E_{-}^{2}\right)+4gE_{+}A_{+}=0,\label{eq:9-1}\\
 &  & \omega_{q}\left(B_{+}^{2}-B_{-}^{2}\right)+\omega_{r}\left(F_{+}^{2}-F_{-}^{2}\right)+4gF_{+}B_{+}=0,\label{eq:10-1}\\
 &  & \omega_{q}\left(A_{+}B_{+}-A_{-}B_{-}\right)+\omega_{r}\left(E_{+}F_{+}-E_{-}F_{-}\right)\nonumber \\
 &  & \qquad\qquad\qquad\qquad+2g\left(A_{+}F_{+}+B_{+}E_{+}\right)=0,\label{eq:11-1}\\
 &  & \omega_{q}A_{-}B_{-}+\omega_{r}E_{-}F_{-}=0,\label{eq:12-1}\\
 &  & A_{-}A_{+}+B_{+}B_{-}=1,\label{eq:13-1}\\
 &  & E_{-}E_{+}+F_{+}F_{-}=1,\label{eq:14-1}\\
 &  & A_{-}E_{+}+B_{-}F_{+}=0,\label{eq:15-1}\\
 &  & A_{+}E_{-}+B_{+}F_{-}=0.\label{eq:16-1}
\end{eqnarray*}

Given that one may solve the Bogoliubov coefficient equtions, we can
determine new frequencies in $\mathcal{H}$:
\begin{eqnarray*}
 &  & \bar{\omega}_{r}=\frac{\omega_{q}}{2}\left(B_{+}^{2}+B_{-}^{2}\right)+\frac{\omega_{r}}{2}\left(F_{+}^{2}+F_{-}^{2}\right)+2gB_{+}F_{+}\label{eq:wr-1}\\
 &  & \bar{\omega}_{q}=\frac{\omega_{q}}{2}\left(A_{+}^{2}+A_{-}^{2}\right)+\frac{\omega_{r}}{2}\left(E_{+}^{2}+E_{-}^{2}\right)+2gA_{+}E_{+}\label{eq:wq-1}
\end{eqnarray*}

One can easily prove that $F_{+}F_{-}=A_{+}A_{-}$, which simplifies
equations and helps to find the following two important equalities: 
\begin{eqnarray*}
E_{+}^{2} & = & \frac{\omega_{r}A_{+}\left(1-A_{-}A_{+}\right)}{\omega_{q}A_{-}},\qquad E_{-}^{2}=\frac{\omega_{q}A_{-}\left(1-A_{-}A_{+}\right)}{\omega_{r}A_{+}}
\end{eqnarray*}

Substituting them in Eq. (\ref{eq:9-1}) we find one equation between
$A_{\pm}$:
\begin{eqnarray}
&& \left[\omega_{q}^{2}A_{-}\left(A_{+}^{3}-A_{-}\right)+\omega_{r}^{2}A_{+}^{2}\left(1-A_{+}A_{-}\right)\right]^{2}\nonumber \\
&& -16\omega_{r}\omega_{q}g^{2}A_{+}^{5}A_{-}\left(1-A_{-}A_{+}\right)  =  0\label{eq:meq1-1}
\end{eqnarray}

This is one of the main equations we need to solve. Another one can
be determined taking some non-trivial steps listed below: We use Eq.
(\ref{eq:11-1}), substitute $B_{\pm}$ from Eqs. (\ref{eq:15-1},\ref{eq:16-1}),
multiply two side in $E_{+}F_{-}^{2}F_{+}$ and simplify it, magically
the final equation is again a second equation that relation $A_{\pm}$:
\begin{eqnarray}
\left(1-A_{-}A_{+}\right)A_{+}A_{-}\left(\frac{\omega_{r}^{2}}{2\omega_{q}}-\frac{\omega_{q}}{2}\right)^{2}\nonumber \\
-\left(2A_{-}A_{+}-1\right)^{2} & = & 0\label{eq:meq2-1}
\end{eqnarray}

Now we solve these two equations together. To do so we first define
$x=A_{+}A_{-}$ and substitute in Eq. (\ref{eq:meq2-1}): $a(1-x)x-(2x-1)^{2}=0$
with $a\equiv\frac{\Delta^{2}\Sigma^{2}}{4g^{2}\omega_{r}\omega_{q}}$ and $\Sigma=\omega_{r}+\omega_{q}$ and $\Delta=\omega_{r}-\omega_{q}$.
Exact real-valued solution is
\[ A_{-}A_{+}=\frac{1}{2}+\frac{1}{2}s,\qquad s^{-1}\equiv\sqrt{1+\frac{16g^{2}\omega_{r}\omega_{q}}{\Delta^{2}\Sigma^{2}}}
\]
and substituting in Eq. (\ref{eq:meq1-1}) determines exact real-valued
$A_{\pm}$:
\begin{eqnarray*}
A_{-} & = & 2^{-\frac{3}{4}}\omega_{q}^{-\frac{1}{2}}\sqrt{1+s}\left(\omega_{q}^{2}+\omega_{r}^{2}-\Delta\Sigma s^{-1}\right)^{\frac{1}{4}}\\
A_{+} & = & 2^{-\frac{1}{4}}\omega_{q}^{\frac{1}{2}}\sqrt{1+s}\left(\omega_{q}^{2}+\omega_{r}^{2}-\Delta\Sigma s^{-1}\right)^{-\frac{1}{4}}\\
E_{-} & = & -2^{-\frac{3}{4}}\omega_{r}^{-\frac{1}{2}}\sqrt{1-s}\left(\omega_{q}^{2}+\omega_{r}^{2}-\Delta\Sigma s^{-1}\right)^{\frac{1}{4}}\\
E_{+} & = & -2^{-\frac{1}{4}}\omega_{r}^{\frac{1}{2}}\sqrt{1-s}\left(\omega_{q}^{2}+\omega_{r}^{2}-\Delta\Sigma s^{-1}\right)^{-\frac{1}{4}}\\
F_{-} & = & 2^{-\frac{3}{4}}\omega_{r}^{-\frac{1}{2}}\sqrt{1+s}\left(\omega_{q}^{2}+\omega_{r}^{2}+\Delta\Sigma s^{-1}\right)^{\frac{1}{4}}\\
F_{+} & = & 2^{-\frac{1}{4}}\omega_{r}^{\frac{1}{2}}\sqrt{1+s}\left(\omega_{q}^{2}+\omega_{r}^{2}+\Delta\Sigma s^{-1}\right)^{-\frac{1}{4}}\\
B_{-} & = & 2^{-\frac{3}{4}}\omega_{q}^{-\frac{1}{2}}\sqrt{1-s}\left(\omega_{q}^{2}+\omega_{r}^{2}+\Delta\Sigma s^{-1}\right)^{\frac{1}{4}}\\
B_{+} & = & 2^{-\frac{1}{4}}\omega_{q}^{\frac{1}{2}}\sqrt{1-s}\left(\omega_{q}^{2}+\omega_{r}^{2}+\Delta\Sigma s^{-1}\right)^{-\frac{1}{4}}
\end{eqnarray*}

In order to find $F_{\pm}$ yet we need to simplify Eq. (\ref{eq:10-1})
by multiplying on both sides on $F_{-}F_{+}$ and rewriting $B_{\pm}$
in terms of $A_{\pm}$, $E_{\pm}$ and $F_{\pm}$ as shown in Eqs.
(\ref{eq:15-1},\ref{eq:16-1}):
\[ \left(\frac{F_{-}}{F_{+}}\right)^{2}=\frac{1}{2}\frac{\omega_{q}^{2}+\omega_{r}^{2}+\Delta\Sigma s^{-1}}{\omega_{r}^{2}}
\]

Defining 
\begin{eqnarray*}
K_{\pm} & \equiv & 2^{-\frac{1}{4}}\left(\omega_{q}^{2}+\omega_{r}^{2}\pm\Delta\Sigma s^{-1}\right)^{\frac{1}{4}}
\end{eqnarray*}

then
\begin{eqnarray*}
 &  & A=\frac{\sqrt{1+s}}{2^{3/2}}\left(\frac{\sqrt{\omega_{q}}}{K_{-}}+\frac{K_{-}}{\sqrt{\omega_{q}}}\right), \\ 
  &  & B=\frac{\sqrt{1-s}}{2^{3/2}}\left(\frac{\sqrt{\omega_{q}}}{K_{+}}+\frac{K_{+}}{\sqrt{\omega_{q}}}\right),\\ 
   & & C=\frac{\sqrt{1+s}}{2^{3/2}}\left(\frac{\sqrt{\omega_{q}}}{K_{-}}-\frac{K_{-}}{\sqrt{\omega_{q}}}\right)\\
& & D=\frac{\sqrt{1-s}}{2^{3/2}}\left(\frac{\sqrt{\omega_{q}}}{K_{+}}-\frac{K_{+}}{\sqrt{\omega_{q}}}\right)\\
 &  & E=\frac{-\sqrt{1-s}}{2^{3/2}}\left(\frac{\sqrt{\omega_{r}}}{K_{-}}+\frac{K_{-}}{\sqrt{\omega_{r}}}\right),\\
 &  & F=\frac{\sqrt{1+s}}{2^{3/2}}\left(\frac{\sqrt{\omega_{r}}}{K_{+}}+\frac{K_{+}}{\sqrt{\omega_{r}}}\right),\\
 & & G=\frac{-\sqrt{1-s}}{2^{3/2}}\left(\frac{\sqrt{\omega_{r}}}{K_{-}}-\frac{K_{-}}{\sqrt{\omega_{r}}}\right)\\
 & & H=\frac{\sqrt{1+s}}{2^{3/2}}\left(\frac{\sqrt{\omega_{r}}}{K_{+}}-\frac{K_{+}}{\sqrt{\omega_{r}}}\right)
\end{eqnarray*}

We can expand the functions in terms of small coupling $g$ to any
order. Below are results up to the fourth order:

Substituting in definition of new frequencies one finds: 
\begin{eqnarray*}
\tilde{\omega}_{r} & = & \left(2s\right)^{-\frac{1}{2}}\sqrt{\left(\omega_{q}^{2}+\omega_{r}^{2}\right)s+\Delta\Sigma}\\
\tilde{\omega}_{q} & = & \left(2s\right)^{-\frac{1}{2}}\sqrt{\left(\omega_{q}^{2}+\omega_{r}^{2}\right)s-\Delta\Sigma}
\end{eqnarray*}

In the weak interaction limit these frequecies turn into Lamb and
Stark shifts. Below we evaluate them up to fourth order: 
\begin{eqnarray*}
\tilde{\omega}_{r} & = & \omega_{r}+\frac{2\omega_{q}g^{2}}{\Delta\Sigma}-\frac{2g^{4}\omega_{q}^{2}\left(5\omega_{r}^{2}-\omega_{q}^{2}\right)}{\omega_{r}\Delta^{3}\Sigma^{3}}+O\left(g^{5}\right)\\
\tilde{\omega}_{q} & = & \omega_{q}-\frac{2g^{2}\omega_{r}}{\Delta\Sigma}-\frac{2g^{4}\omega_{r}^{2}\left(\omega_{r}^{2}-5\omega_{q}^{2}\right)}{\omega_{q}\Delta^{3}\Sigma{}^{3}}+O\left(g^{5}\right)
\end{eqnarray*}

Anharmonicity can be easily derived using the following relation:
\begin{eqnarray*}
\left(a-a^{\dagger}\right)^{4} & = & 6\left(A-C\right)^{4}\left(\left(\alpha^{\dagger}\alpha\right)^{2}+\alpha^{\dagger}\alpha\right)\\
 &  & +6\left(B-D\right)^{4}\left(\left(\beta^{\dagger}\beta\right)^{2}+\beta^{\dagger}\beta\right)\\
 &  & +12\left(A-C\right)^{2}\left(B-D\right)^{2}\left(2\alpha^{\dagger}\alpha\beta^{\dagger}\beta+\alpha^{\dagger}\alpha+\beta^{\dagger}\beta\right)
\end{eqnarray*}

\end{document}